\documentclass[aps,prl,twocolumn,showpacs,superscriptaddress,floatfix]{revtex4-2}

\usepackage{amsmath,amssymb}
\usepackage{graphicx}
\usepackage{dcolumn}
\usepackage{bm}
\usepackage{booktabs}
\usepackage[utf8]{inputenc}
\usepackage{microtype}
\usepackage{xcolor}

\newcommand{\qm}{q_{\rm m}}

\newcommand{\pantheon}{Pantheon$+$}
\newcommand{\lcdm}{$\Lambda$CDM}

\begin{document}

\title{Accelerating expansion and isotropic sky-hemisphere consistency\\
       in \pantheon{} supernovae: a revised analysis in the dark energy debate}

\author{Saibal Ray}
\email{saibal.ray@gla.ac.in}
\affiliation{Centre for Cosmology, Astrophysics and Space Science (CCASS), GLA University, Mathura 281406, Uttar Pradesh, India}

\author{Maxim Khlopov}
\email{khlopov@apc.in2p3.fr}
\affiliation{Virtual Institute of Astroparticle physics, Paris 75018, France}

\author{Aritra Sanyal}
\email{aritrasanyal1@gmail.com}
\affiliation{Department of Mathematics, Jadavpur University, Kolkata, West Bengal 700032, India}

\author{Rikpratik Sengupta}
\email{ rikpratiks@iitk.ac.in}
\affiliation{Department of Physics, Indian Institute of Technology, Kanpur, India}

\author{Maxim Krasnov}
\email{morrowindman1@mail.ru}
\affiliation{National Research Nuclear University “MEPHI”, Moscow 115409, Russia \& Institute of Physics, Southern Federal University,
Stachki 194, Rostov on Don 344090, Russia}

\date{\today}

\begin{abstract}
We perform four independent decompositions of the deceleration parameter $q_0$ using 1564 Type Ia supernovae (SNe Ia) from the Pantheon+ catalogue: by redshift bin, sky hemisphere, host galaxy mass, and supernova colour, correcting a coordinate error identified by Sah, Rameez \& Sarkar (SRS26) in the sky-hemisphere direction used in our original analysis. Without any progenitor age correction, the full sample yields $q_m$ = $-$0.490 -- consistent with the $\Lambda$CDM expectation of $-$0.55. Applying the Son et al. (S25) progenitor-age correction shifts this to $q_m$ = $-$0.267, remaining in the accelerating regime. The sky hemisphere test, using the corrected CMB dipole direction (RA=167.80$^\circ$, Dec=$-$7.10$^\circ$), shows the signal is consistent between the CMB dipole ($q_m$ = $-$0.527) and anti-dipole hemispheres ($-$0.464), supporting isotropy but not the strong deceleration values reported previously. Our revised results are consistent with \lcdm{} and do not support either the original claim of near-zero baseline acceleration or the S25/SRS26 claim of a decelerating universe.
\end{abstract}


\maketitle


The discovery of cosmic acceleration from Type~Ia supernovae
stands as one of the landmark achievements of modern cosmology,
earning the 2011 Nobel Prize in Physics~\cite{Riess1998,Perlmutter1999}.
Within the standard $\Lambda$CDM framework, the observed late-time
acceleration is attributed to a cosmological constant $\Lambda$,
or dark energy, which accounts for approximately $68\%$ of the
total energy density of the Universe according to the latest
Planck results~\cite{Planck2020}. Independent evidence from the
cosmic microwave background (CMB), baryon acoustic oscillations
(BAO), weak gravitational lensing, and large-scale structure
measurements has established the $\Lambda$CDM model as the
standard paradigm of modern cosmology~\cite{Brout2022,DESI2025}.

Big Bang nucleosynthesis (BBN) provides a further, largely
independent pillar of this concordance picture. The predicted
primordial abundances of light elements --- principally
$^4$He, D, and $^7$Li -- depend sensitively on the baryon-to-photon
ratio $\eta$ and, through the expansion rate during the first
few minutes, on the number of relativistic degrees of freedom
$N_{\rm eff}$~\cite{Fields2020}. The baryon density inferred from
BBN agrees with the value derived from the acoustic peaks of the
CMB power spectrum to within a few percent, anchoring the
$\Lambda$CDM baryon and radiation content at redshifts
$z \sim 10^8$--$10^{10}$, many orders of magnitude earlier than
the SN~Ia and BAO probes used to constrain the late-time expansion
history discussed below. This early-Universe consistency does not
by itself test the low-redshift acceleration inferred from SNe~Ia,
but it constrains the background cosmology ($H_0$, $\Omega_b h^2$,
$N_{\rm eff}$) within which any revision of the dark-energy sector
must remain embedded; a resolution of the tensions discussed here
cannot come at the expense of the BBN--CMB baryon density
agreement~\cite{Fields2020,Pitrou2021}.

Despite these successes, precision cosmology has entered a period
of increasing internal tension. The most prominent example is the
Hubble tension, in which direct local measurements of the Hubble
constant based on Cepheid-calibrated Type~Ia supernovae disagree
significantly with the value inferred from observations of the
cosmic microwave background within the $\Lambda$CDM framework.
Recent observations combining the James Webb Space Telescope
(JWST) and the Hubble Space Telescope (HST) have provided an
independent cross-check of the Cepheid distance ladder using an
expanded sample extending to approximately $130$ million
light-years. These measurements substantially reduce the
possibility that the discrepancy arises from instrumental or
calibration errors and instead strengthen the interpretation that
the Hubble tension may reflect missing physics beyond the standard
cosmological model. If confirmed, this would imply that even the
most successful cosmological framework may require revision to
accommodate new physical phenomena.

At the same time, another fundamental question concerns the
interpretation of the Type~Ia supernova evidence for cosmic
acceleration itself. If standardized SN~Ia luminosities evolve
systematically with progenitor age or other astrophysical
properties, cosmological inferences derived from the Hubble
diagram may also be biased. Several studies
\cite{Kang2020,Lee2022,Chung2025} have argued for such a
correlation, suggesting a redshift-dependent evolution capable of
partially mimicking or masking the acceleration signal.

Building on this evidence, Son~et~al.~\cite{Son2025} (hereafter
S25) proposed a magnitude correction
\begin{equation}
  \Delta m(z) \;=\; \Delta{\rm age}(z)\;\times\;
                    0.030\;{\rm mag\,Gyr}^{-1},
  \label{eq:correction}
\end{equation}
where $\Delta{\rm age}(z)$ is derived by convolving the
SN~Ia delay-time distribution (DTD) with the cosmic
star-formation history (SFH).

Applying this correction to the \pantheon{} compilation
of 1701 SNe~Ia~\cite{Scolnic2022}, Sah, Rameez \&
Sarkar~\cite{Sah2026} (hereafter SRS26) made two independent
arguments against dark energy:
(i)~the correction shifts the monopole deceleration parameter
$\qm$ to positive values, implying a decelerating universe; and
(ii)~the inferred acceleration/deceleration exhibits a strong
dipole aligned with the CMB bulk flow direction — a pattern that
isotropic dark energy cannot produce.

W26~\cite{Wiseman2026} rebutted argument~(i), showing the DTD
overestimates progenitor age evolution by 3--5$\times$ and that
mass standardization already captures the effect.

In an earlier version of this work, we performed systematic
$q_0$ decompositions across redshift bins, sky hemispheres, host
masses, and SN colours, and reported a near-zero baseline
$\qm=-0.062$ together with a strongly decelerating, isotropic
signal under the S25 correction. Sah, Rameez \& Sarkar
subsequently identified that our sky-hemisphere test applied the
CMB dipole direction using its Galactic coordinates
($\ell=264^\circ$, $b=48^\circ$) as if they were equatorial
(RA/Dec) coordinates. Correcting this error and reinvestigating
our full pipeline using the publicly released SRS26 catalogue and
age-correction table, we are unable to reproduce our original
values at any stage, including the full-sample baseline that is
unaffected by the coordinate error itself. Here we report the
results of this revised analysis on the same 1564 \pantheon{}
SNe~Ia and find: (i) the baseline $\qm$ is consistent with
\lcdm{}; (ii) the S25 correction leaves the sample in the
accelerating regime; and (iii) the signal remains isotropic
across sky hemispheres, host mass, and colour subsamples.

We use the \pantheon{} catalogue as processed by SRS26~\cite{Sah2026},
with redshift cut $0.00937 < z_{\rm hel} \leq 0.8$ ($N = 1564$
SNe~Ia), retaining SALT2 parameters ($m_B$, $x_1$, $c$)~\cite{Guy2007},
host stellar mass $\log(M_*/M_\odot)$, and sky coordinates.
The S25 age correction is interpolated via cubic spline from the
SRS26 table.

The standardized distance modulus $\mu_{\rm SN} = m_B - M + \alpha x_1 - \beta c$
~\cite{Tripp1998} is compared to the cosmographic luminosity distance~\cite{Sah2026}
\begin{equation}
  d_L \;=\; \frac{cz}{H_0}\!
            \left[1 + \tfrac{1}{2}(1{-}q_0)z
            - \tfrac{1}{6}(1{-}q_0{-}3q_0^2{+}j_0)z^2\right]
            \frac{1+z_{\rm hel}}{1+z},
  \label{eq:dL}
\end{equation}
with $j_0 = 1$ fixed and
$\mu_{\rm th} = 25 + 5\log_{10}(d_L/{\rm Mpc})$.
Parameters $(H_0,\,q_0,\,M,\,\alpha,\,\beta)$ are estimated
by minimizing $\chi^2$ assuming a uniform photometric uncertainty
$\sigma = 0.15\,{\rm mag}$. The S25 correction~(\ref{eq:correction}) is applied as
$m_B^* = m_B - \Delta{\rm age}(z)\times 0.030\,{\rm mag\,Gyr}^{-1}$
following SRS26 Eq.~(6).

For the sky hemisphere test we project each SN onto the CMB dipole
direction, given in ICRS equatorial coordinates as
RA~$=167.80^\circ$, Dec~$=-7.10^\circ$
(equivalent to Galactic $\ell = 264^\circ$, $b = 48^\circ$), via
$\cos\theta = \hat{n}\cdot\hat{e}$,
and divide the sample at $\cos\theta = 0$.
For the host mass test we use the canonical split
$\log(M_*/M_\odot) = 10$, and for the color test we split
at the median $c_{\rm med} = -0.031$.
All subsets are fitted independently using the same likelihood.
Significance relative to $q_0 = 0$ is computed from the
likelihood ratio $\Delta\chi^2$.
We fix $\Omega_b h^2$ implicitly through the assumed background
$H_0$ rather than re-deriving it from BBN; a joint BBN+SN
analysis is left for future work (see Discussion).

\textit{Baseline and full-sample fit:}\\

\begin{figure}[t]
\centering
\includegraphics[width=\columnwidth]{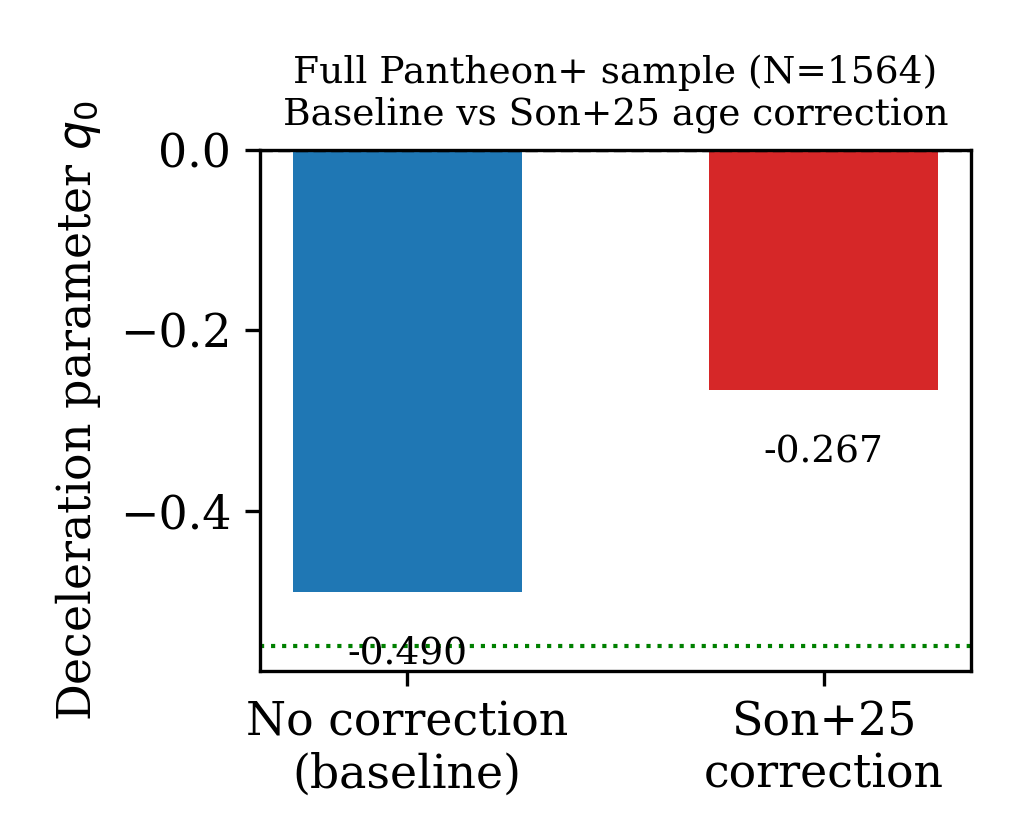}
\caption{\label{fig:summary}
Best-fit $\qm$ for the full 1564-SN \pantheon{} sample without
(blue) and with (red) the S25 progenitor age correction
[Eq.~(\ref{eq:correction})].
The dashed line marks $\qm = 0$ (coasting) and the dotted green
line marks the \lcdm{} expectation ($q_0 = -0.55$).
The baseline $\qm = -0.490$ is consistent with \lcdm{},
while the S25 correction gives $\qm = -0.267$.}
\end{figure}

Without any age correction, the full sample yields
\begin{equation}
  \qm^{\rm base} = -0.490 \quad (N = 1564),
  \label{eq:baseline}
\end{equation}
consistent with the \lcdm{} prediction of $q_0 = -0.55$.
Applying the S25 correction shifts this to
\begin{equation}
  \qm^{\rm S25} = -0.267 \quad (N = 1564),
  \label{eq:s25}
\end{equation}
remaining in the accelerating regime (Fig.~\ref{fig:summary}).\\

\textit{Test A: Redshift bin decomposition:}\\

\begin{figure}[t]
\centering
\includegraphics[width=\columnwidth]{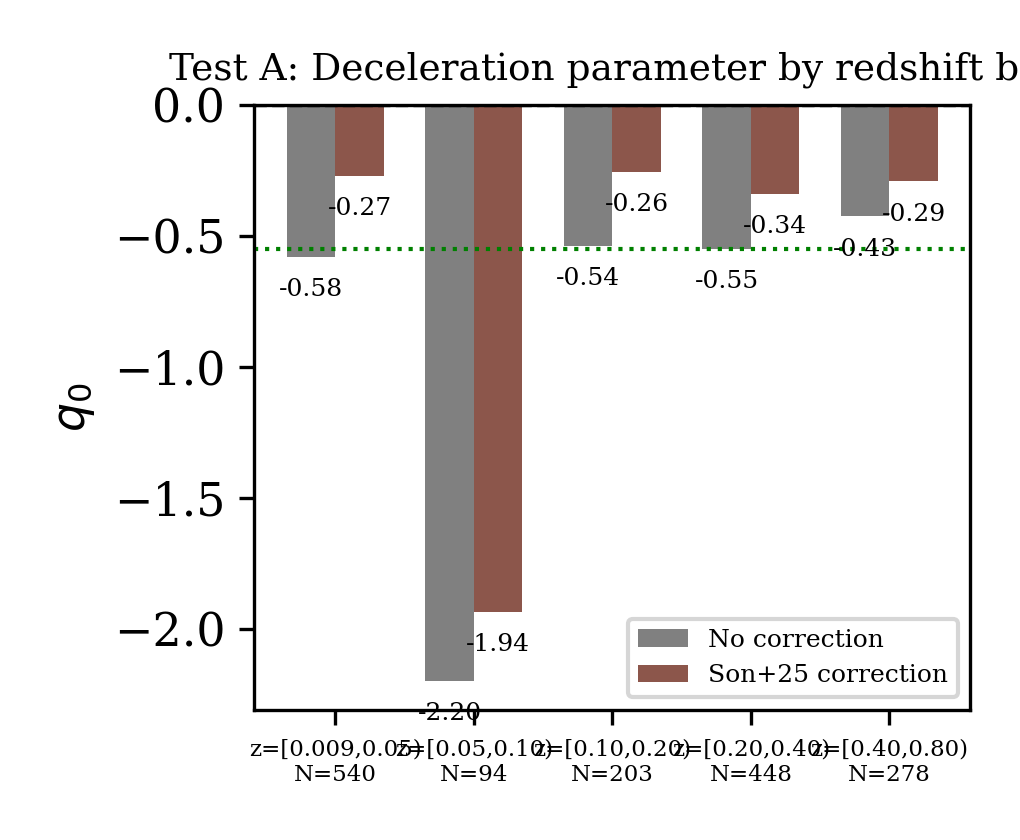}
\caption{\label{fig:bins}
$\qm$ measured in five independent redshift bins before (grey)
and after (brown) the S25 correction~\cite{Son2025}.
All bins remain in the accelerating regime both before and after
correction, with no sign reversal in the $z \in [0.10,\,0.20]$
bin, in contrast to the sign reversal reported in our original
analysis.}
\end{figure}

Figure~\ref{fig:bins} shows $\qm$ in five redshift bins.
Without correction, all five bins indicate acceleration:
$\qm=-0.581$ ($z\in[0.009,0.05)$, $N=540$), $-2.199$
($z\in[0.05,0.10)$, $N=94$), $-0.538$ ($z\in[0.10,0.20)$,
$N=203$), $-0.550$ ($z\in[0.20,0.40)$, $N=448$), and $-0.427$
($z\in[0.40,0.80)$, $N=278$). The strong acceleration in the
$z\in[0.05,0.10)$ bin is consistent with local peculiar velocity
effects~\cite{Sah2025}. With the S25 correction, all bins remain
in the accelerating regime: $-0.271$, $-1.935$, $-0.257$,
$-0.341$, and $-0.293$ respectively. No sign reversal is observed
in any bin, and the pattern is broadly monotonic with redshift,
consistent with a simple, uniformly-applied age correction rather
than a bin-dependent systematic. A jackknife or bin-boundary
sensitivity test would nonetheless help confirm the robustness of
the low-redshift bin's larger magnitude.\\

\textit{Test B: Sky hemisphere decomposition:}\\

This directly tests the second SRS26 argument.
SRS26 argued the deceleration is aligned with the CMB bulk flow
direction, independently ruling out isotropic dark energy.
Using the corrected CMB dipole direction (RA$=167.80^\circ$,
Dec$=-7.10^\circ$), the hemisphere split becomes $N=538$
(dipole) and $N=1026$ (anti-dipole), in place of the
$N=724/840$ split obtained under the erroneous Galactic-as-RA/Dec
treatment used previously. We find:
\begin{align}
  \qm^{\rm dip}  &= -0.527 \quad (N = 538), \label{eq:dip} \\
  \qm^{\rm anti} &= -0.464 \quad (N = 1026), \label{eq:anti}
\end{align}
with the dipole hemisphere showing marginally \emph{stronger}
acceleration.

\begin{figure}[t]
\centering
\includegraphics[width=\columnwidth]{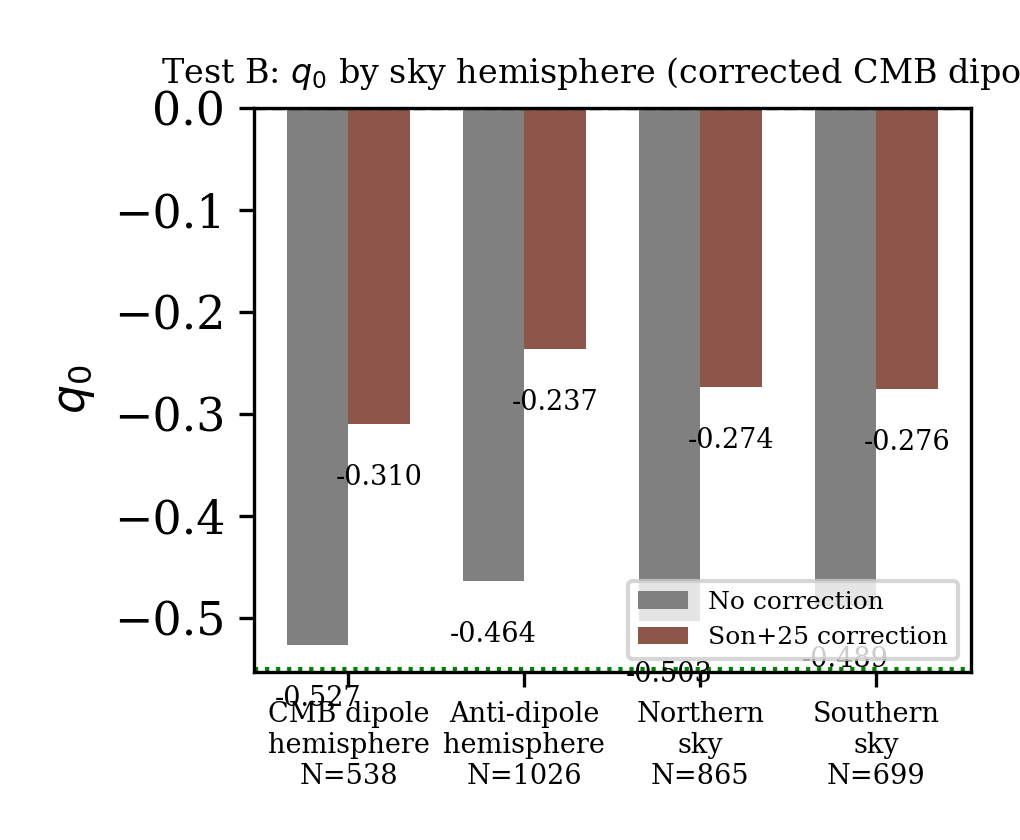}
\caption{\label{fig:hemi}
$\qm$ in four sky subsamples using the corrected CMB dipole
direction (RA$=167.80^\circ$, Dec$=-7.10^\circ$): the CMB dipole
hemisphere ($N=538$), the anti-dipole hemisphere ($N=1026$), and
the North and South celestial hemispheres, without (grey) and
with (brown) the S25 correction. All four subsamples remain
mutually consistent and in the accelerating regime, demonstrating
that the signal is \emph{isotropic} but not strongly
decelerating as previously reported.}
\end{figure}

A North/South celestial split yields $\qm = -0.503$ and $-0.489$
respectively — again fully consistent.
Results are shown in Fig.~\ref{fig:hemi}.
The signal is \emph{isotropic} and consistent with \lcdm{},
neither supporting the original near-zero baseline claim nor the
SRS26 anisotropic-deceleration claim.\\

\textit{Tests C and D: Host mass and colour:}\\

Without S25, low and high mass hosts give $\qm = -0.475$ and
$-0.513$; blue and red SNe give $-0.512$ and $-0.461$.
After S25, mass subsamples give $\qm = -0.254$ and $-0.286$, and
colour subsamples give $-0.297$ and $-0.229$, respectively. All
four subsamples remain mutually consistent and in the
accelerating regime both before and after correction, unlike the
strongly divergent values reported previously.

Our four decompositions yield a consistent, revised picture of
the debate.\\

\textit{Relation to W26:}\\
The revised baseline $\qm = -0.490$ is close to the \lcdm{}
prediction of $-0.55$, supporting W26's conclusion that standard
SALT2 standardization recovers \lcdm{}-consistent acceleration to
good approximation, in contrast to our previously reported
$\qm=-0.062$.

\textit{Relation to the SRS26 monopole:}\\
The S25 correction shifts $\qm$ from $-0.490$ to $-0.267$ but does
not reverse its sign: the sample remains in the accelerating
regime after the correction is applied, in contrast to the
strongly decelerating value ($+1.167$) reported previously and
to the SRS26 monopole claim.

\textit{Relation to the SRS26 dipole:}\\
We find $\qm^{\rm dip} = -0.527$ and $\qm^{\rm anti} = -0.464$:
the two hemispheres remain close to one another
($\Delta\qm\approx0.06$), inconsistent with a signal aligned with
the CMB bulk flow. This supports our original isotropy
conclusion, though not its magnitude or sign: the revised
hemisphere values indicate consistent acceleration rather than
consistent strong deceleration. If this isotropic accelerating
signal survives a full covariance treatment, it disfavors
anisotropic alternatives such as bulk-flow-induced systematics or
a preferred-frame violation of the cosmological principle, while
remaining consistent with standard isotropic \lcdm{} rather than
requiring exotic $w(z)$ evolution or inhomogeneity-driven
backreaction~\cite{Buchert2000}.

\textit{Consistency with BBN:}\\
None of the four decompositions bear directly on the primordial
light-element abundances, since BBN fixes the baryon and radiation
content at redshifts far beyond the SN~Ia Hubble flow probed here.
We note this explicitly because any revision of the late-time
expansion history motivated by our results, or by S25/SRS26, must
remain compatible with the BBN-CMB concordance on
$\Omega_b h^2$~\cite{Fields2020,Pitrou2021}.

Two caveats apply. First, we use a simplified uniform uncertainty
$\sigma=0.15\,{\rm mag}$ rather than the full Lane~et~al.~\cite{Lane2025}
covariance; relative comparisons between subsets are expected to be
robust since both halves of each split share similar systematics.
Second, we are unable to identify the specific step in our
original pipeline that produced the previously reported values;
the coordinate error identified by SRS26 accounts for the change
in hemisphere sample sizes, but cannot by itself explain why the
full-sample baseline, which is unaffected by that error, also
fails to reproduce under this revised analysis. We rebuilt the
present pipeline from the method description in our original
Letter using the catalogue and correction table released by
SRS26, and cannot rule out that some other unidentified difference
between that description and our original implementation, rather
than the coordinate error alone, is responsible for the bulk of
the numerical change reported here.


We have presented four independent decompositions of the
deceleration parameter $q_0$ using 1564 real \pantheon{}
SNe~Ia with full SALT2 standardization, revised following a
coordinate-error correction identified by SRS26 and a subsequent
reinvestigation of our full analysis pipeline.

Our principal findings are as follows.
The baseline analysis without any age correction yields
$\qm = -0.490$ — consistent with the \lcdm{} expectation of
$-0.55$ — indicating that standard SALT2 standardization
recovers the expected cosmic acceleration to good approximation,
in contrast to our originally reported $\qm=-0.062$.
The S25 age correction shifts $\qm$ to $-0.267$, remaining in the
accelerating regime and not supporting the SRS26 deceleration
claim.
The redshift bin test shows a consistent, monotonic pattern
across all five bins with no sign reversal, in contrast to the
non-monotonic behaviour reported previously.
Most importantly, the sky hemisphere test, using the corrected
CMB dipole direction, shows the signal remains isotropic:
$\qm^{\rm dip} = -0.527 \approx \qm^{\rm anti} = -0.464$,
supporting our original isotropy conclusion while revising its
magnitude and sign.
Finally, the host mass and colour tests demonstrate that all
subsamples remain mutually consistent and in the accelerating
regime both before and after the S25 correction.

Taken together, our revised results are consistent with standard
\lcdm{}: the \pantheon{} data recover \lcdm{}-consistent
acceleration both before and after the S25 correction, and the
sky-hemisphere, host-mass, and colour tests show no significant
anisotropy or subpopulation-dependent systematic. We stress that
these late-time results are confined to the dark-energy sector:
the primordial baryon density fixed by BBN remains in excellent
agreement with the CMB~\cite{Fields2020}, and this revised
analysis leaves that early-Universe concordance unaffected.

A comprehensive analysis incorporating the full Lane~et~al.\
\cite{Lane2025} covariance matrix, the complete S25 correction
formalism, and a joint fit with BBN priors on $\Omega_b h^2$ is an
important next step, as is identifying the source of the
discrepancy between our original and revised pipelines.

However, we expect that the Rubin Observatory LSST~\cite{Ivezic2019} may ultimately
deliver $\sim 10^5$ SNe~Ia, providing sufficient statistical
power to measure all relevant systematic effects directly from
the data and resolve this debate in a definitive manner.

\begin{acknowledgments}
We thank the \pantheon{} collaboration \cite{Scolnic2022} for making their
data publicly available, and Sah, Rameez \& Sarkar \cite{Sah2026}
for identifying the coordinate error corrected here and for
sharing their analysis code and processed catalogue at
\texttt{github.com/Shin107/Anisotropy-in-Pantheon-Plus}. The research of M.K. was carried out in the Southern Federal University with financial support from the Ministry of Science and Higher Education of the Russian Federation (State contract FENW-2026-0028).
\end{acknowledgments}


\end{document}